\pdfoutput=1

\documentclass[11pt,twoside,a4paper,cmspaper,final,collab]{cms-tdr}
\def\svnVersion{493393P}\def\cmsCernNoTag{CERN-EP-2019-014}\def\cmsCernDate{\today}\def\cmsMessage{Published in Physical Review Letters as \href{http://dx.doi.org/10.1103/PhysRevLett.122.132001}{\doi{10.1103/PhysRevLett.122.132001.}}}

\begin{document}\cmsNoteHeader{BPH-18-007}

\hyphenation{had-ron-i-za-tion}
\hyphenation{cal-or-i-me-ter}
\hyphenation{de-vices}
\RCS$HeadURL: svn+ssh://svn.cern.ch/reps/tdr2/papers/BPH-18-007/trunk/BPH-18-007.tex $
\RCS$Id: BPH-18-007.tex 490327 2019-02-27 09:59:48Z carlosl $

\newlength\cmsFigWidth
\ifthenelse{\boolean{cms@external}}{\setlength\cmsFigWidth{0.98\columnwidth}}{\setlength\cmsFigWidth{0.65\textwidth}}

\ifthenelse{\boolean{cms@external}}{
\newcommand{\Bc}{\ensuremath{\BC^{+}}\xspace}
\newcommand{\BcStar}{\ensuremath{\BC^{*+}}\xspace}
}{
\newcommand{\Bc}{\ensuremath{\BC^{\!\!\!+}}\xspace}
\newcommand{\BcStar}{\ensuremath{\BC^{\!\!\!*+}}\xspace}
}
\newcommand{\BcPrime}{\ensuremath{\Bc\cmsSymbolFace{(2S)}}\xspace}
\newcommand{\BcPrimeStar}{\ensuremath{\BcStar\cmsSymbolFace{(2S)}}\xspace}

\newcommand{\Bcpipi}{\ensuremath{\Bc \, \Pgpp \Pgpm}\xspace}
\newcommand{\BctoJpsipi}{\ensuremath{\Bc \! \to \! \cPJgy \, \Pgpp}\xspace}
\newcommand{\BctoJpsiK}{\ensuremath{\Bc \! \to \! \cPJgy \, \PK^{+}}\xspace}
\newcommand{\BctoJpsiX}{\ensuremath{\Bc \! \to \! \cPJgy \, \Pgpp \, X}\xspace}

\newcommand{\barbc}{\ensuremath{\cPaqb\cPqc}\xspace}

\newcommand{\Jpsipi}{\ensuremath{\cPJgy \, \Pgpp}\xspace}

\cmsNoteHeader{BPH-18-007}

\title{Observation of two excited \texorpdfstring{\Bc states
and measurement of the \BcPrime mass
in $\Pp\Pp$ collisions at $\sqrt{s} = 13\TeV$}{Bc+ states
and measurement of the Bc+(2S) mass
in pp collisions at sqrt(s) = 13 TeV}}

\abstract{
Signals consistent with the
\BcPrime and \BcPrimeStar states are observed in proton-proton collisions at $\sqrt{s} = 13\TeV$,
in an event sample corresponding to an integrated luminosity of 143\fbinv,
collected by the CMS experiment during the 2015--2018 LHC running periods.
These excited $\barbc$ states are observed in the \Bcpipi invariant mass spectrum,
with the ground state \Bc reconstructed through its decay to \Jpsipi.
The two states are reconstructed as two well-resolved peaks,
separated in mass by $29.1 \pm 1.5\stat \pm 0.7\syst$\MeV.
The observation of two peaks, rather than one, 
is established with a significance exceeding five standard deviations.
The mass of the \BcPrime meson is measured to be
$6871.0 \pm 1.2 \stat \pm 0.8 \syst \pm 0.8 \, (\Bc)$\MeV,
where the last term corresponds to the uncertainty in the world-average \Bc mass.}

\hypersetup{%
pdfauthor={CMS Collaboration},%
pdftitle={Observation of the Bc(2S) and Bc*(2S) states and measurement of the Bc(2S) mass
in pp collisions at 13 TeV},%
pdfsubject={CMS},%
pdfkeywords={CMS, physics, B mesons, quarkonium, spectroscopy}}

\maketitle

\begin{figure}[ht]
\centering
\includegraphics[width=0.38\textwidth]{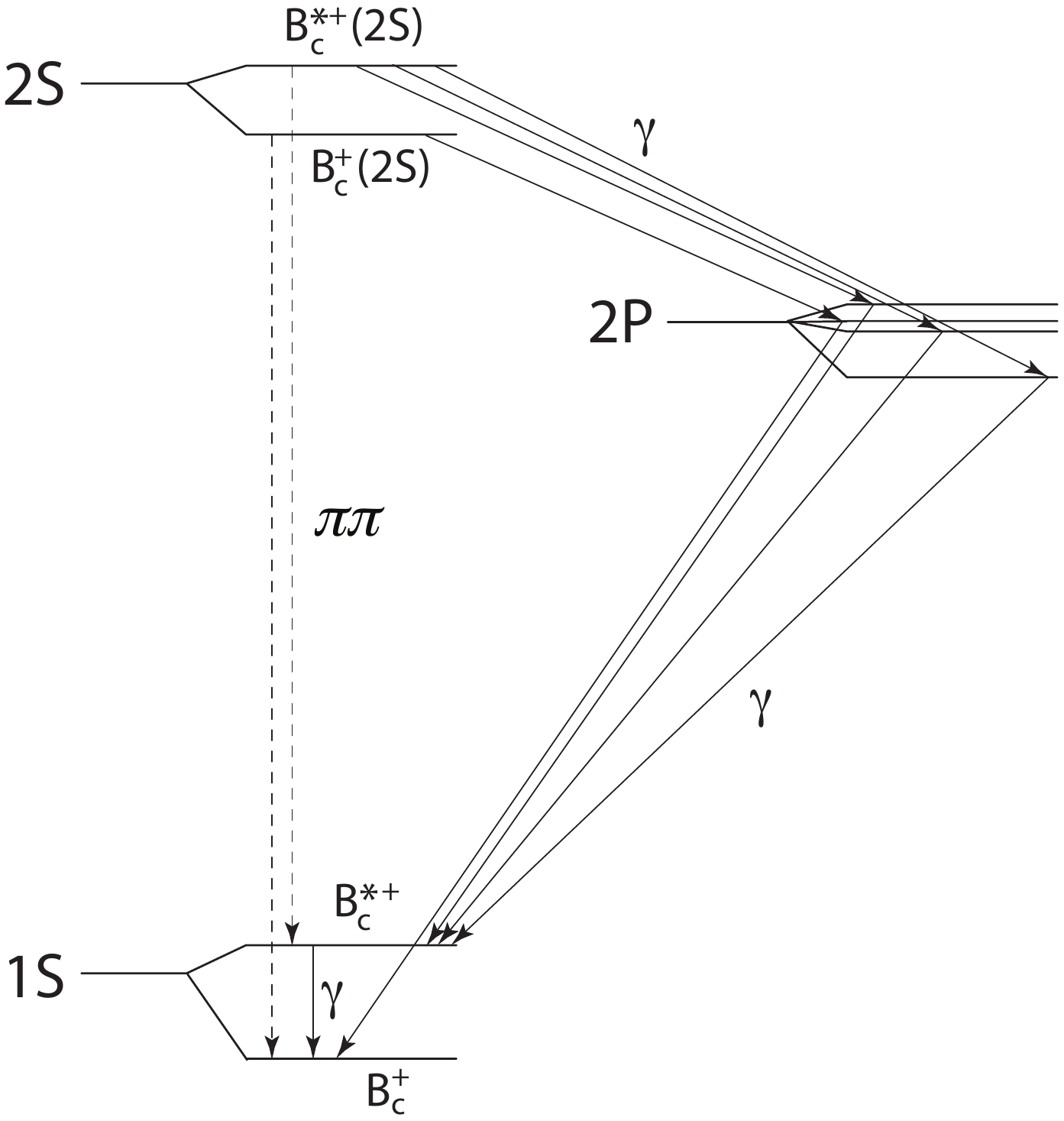}
\caption{Transitions between the lightest \BC states,
with solid and dashed lines indicating the emission of photons and pion pairs, respectively~\cite{PhysRevD.49.5845}.}
\label{fig:figure1}
\end{figure}

The \BC family consists of charged mesons composed of a beauty quark and a charm antiquark (or vice-versa).
The ground state was discovered in 1998 by the CDF Collaboration~\cite{Abe:1998wi}.
The spectrum of this
heavy quarkonium
family is predicted to be very
populated~\cite{PhysRevD.49.5845,Kiselev:1994rc,Zeng:1994vj,Gupta:1995ps,Davies:1996gi,Fulcher:1998ka,Ebert:2002pp,PhysRevD.70.054017,Berezhnoy:2013sla,Gregory:2009hq,Dowdall:2012ab,Mathur:2018epb},
but spectroscopic observations and measurements of production properties remain scarce.
Indeed, their production yields are significantly smaller than those of the charmonium and bottomonium states,
the $\barbc$ production cross sections being proportional to the fourth power of the strong coupling constant, $\alpS^{4}$
(since two pairs of heavy quarks need to be produced).
While the masses and sizes of these beauty-charm quark-antiquark pairs
place them between the charmonium and bottomonium systems,
so that many properties can be theoretically inferred by interpolation of existing knowledge,
the unequal quark masses and velocities could lead to more complex dynamics,
where some (nonrelativistic) approximations might break down.
Since the $\barbc$ mesons cannot annihilate into gluons,
the excited states decay to the ground state via the cascade emission of photons
or pion pairs,
leading to total widths that are less than a few hundred\,\keV.
Figure~\ref{fig:figure1} shows the transitions between the lightest \BC states.

The high collision energies and integrated luminosities provided by the LHC have opened the way for a series of new measurements.
The ATLAS Collaboration observed a state with a mass ($6842 \pm 4\stat \pm 5\syst$\MeV)
consistent with the values predicted for the \BcPrime,
using data collected at 7 and 8\TeV~\cite{Aad:2014laa},
while the LHCb Collaboration reported that their 8\TeV data sample did not show any significant sign of the
\BcPrime or \BcPrimeStar states~\cite{Aaij:2017lpg}.
The peak observed by ATLAS could be the superposition of the \BcPrime and \BcPrimeStar states,
too closely spaced with respect to the resolution of the measurement.
The mass difference between the \BcStar and \Bc hyperfine partners is predicted to be around 55\MeV,
while the corresponding difference between the \BcPrimeStar and \BcPrime masses should be
around 35\MeV~\cite{Gregory:2009hq,Dowdall:2012ab,Mathur:2018epb}.

While the \BcPrime decays directly to $\Bc \, \Pgpp \Pgpm$,
the \BcPrimeStar is expected to decay predominantly to $\BcStar \, \Pgp^+ \Pgp^-$, followed by the $\BcStar \! \to \! \Bc \, \gamma$ decay.
The emitted photon has a very low energy and its detection is very challenging,
so that the \BcPrimeStar peak should be seen in the $\Bc \, \Pgpp \Pgpm$ mass spectrum
at the mass $M(\BcPrime) - \Delta M$, where $\Delta M \equiv [ M(\BcStar) - M(\Bc) ] - [ M(\BcPrimeStar) - M(\BcPrime)]$.
If the $\Delta M$ value is larger than the experimental resolution,
the $\Bc \, \Pgp^+ \Pgp^-$ invariant mass distribution will show a two-peak structure.
Since $M(\BcStar) - M(\Bc)$ is predicted to be larger than $M(\BcPrimeStar) - M(\BcPrime)$,
the \BcPrimeStar state will be the lower mass peak.

This Letter reports the observation of well-resolved
signals consistent with the \BcPrime and \BcPrimeStar states,
as well as the first measurement of the \BcPrime mass.
Although strictly speaking we should refer to these two signals as
\BcPrime and \BcPrimeStar candidates, in the remainder of the
text we will skip the word candidates for improved readability.
The result is based on the analysis of proton-proton data samples collected by the CMS experiment at a center-of-mass energy of 13\TeV,
in 2015, 2016, 2017, and 2018 (the full LHC Run~2),
corresponding to integrated luminosities of 2.8, 36.1, 42.1, and 61.6\fbinv, respectively.

The central feature of the CMS apparatus is a superconducting solenoid of 6\unit{m} internal diameter,
providing a magnetic field of 3.8\unit{T}.
Within the solenoid volume are a silicon pixel and strip tracker, a lead tungstate crystal electromagnetic calorimeter,
and a brass and scintillator hadron calorimeter, each composed of a barrel and two endcap sections.
Forward calorimeters extend the pseudorapidity coverage provided by the barrel and endcap detectors.
Muons are detected in gas-ionization chambers embedded in the steel flux-return yoke outside the solenoid.
A more detailed description of the CMS detector, together with a definition of the coordinate system used and
the relevant kinematic variables, can be found in Ref.~\cite{bib:CMSJINST}.

The event samples used in this analysis were collected with a two-level trigger system~\cite{bib:CMStrigger}.
The first level consists of custom hardware processors and uses information from the muon system to select events with two muons.
The high-level trigger requires
two oppositely charged muons with pseudorapidity $\abs{\eta} < 2.5$ and transverse momentum $\pt > 4$\GeV,
a distance of closest approach between the two muons smaller than 0.5\cm,
a dimuon vertex fit $\chi^2$ probability larger than 10\%,
a dimuon invariant mass in the range 2.9--3.3\GeV,
and a distance between the dimuon vertex and the beam axis larger than three times its uncertainty.
In addition, the dimuon \pt must be aligned with the transverse displacement vector:
$\cos\theta > 0.9$, where ${\cos\theta} = {\vec{L}_{xy}\cdot \vec{\pt} / (L_{xy} \, \pt)}$,
with $\vec{L}_{xy}$ representing the transverse decay displacement vector of the dimuon.
Finally, there must exist a third track in the event compatible with being produced at the dimuon vertex.
The offline reconstruction requires
two oppositely charged muons matching those that triggered the detector readout,
with some requirements being stricter than at the trigger level,
such as $\abs{\eta} < 2.4$ and $\cos\theta > 0.98$.
The muons must fulfill the ``soft muon identification" requirements~\cite{bib:softmuon2}
and be close to each other in angular space:
$\sqrt{\smash[b]{(\Delta\eta)^2 + (\Delta\phi)^2}} < 1.2$,
where $\Delta\eta$ and $\Delta\phi$ are
differences in pseudorapidity and azimuthal angle, respectively,
between the directions of the two muons.

Several simulated samples were used in the analysis.
The \Bc, \BcPrime, and \BcPrimeStar signal samples are generated with the \textsc{bcvegpy}~2.2~\cite{Bcveg} Monte Carlo generator,
interfaced with the \PYTHIA~8.230 package~\cite{bib:Pythia} to simulate the hadronization step,
and with \EVTGEN~1.6.0~\cite{bib:EvtGen} for the decays.
Final-state radiation is modeled with \PHOTOS~3.61~\cite{bib:PHOTOS2}.
The generated events are then processed through a detailed simulation of the CMS detector, based on the \GEANTfour package~\cite{geant4},
using the same trigger and reconstruction algorithms as used for the collision data.
The simulated events include multiple proton-proton interactions in the same or nearby beam crossings,
with a distribution matching the measured one.
Charge-conjugated states are implied throughout this Letter.

All the physics objects used in this analysis, including the muon tracks, 
must pass high-purity track quality requirements~\cite{TRK-11-001}. 
The \Bc candidates are reconstructed by combining the dimuon with a track,
assumed to be a pion.
This track must have at least one hit in the pixel layers, at least five hits in the tracker (pixel and strip layers),
an impact parameter in the transverse plane larger than two times its uncertainty,
$\abs{\eta} < 2.4$, and $\pt > 3.5$\GeV.
The \Bc candidate is obtained by performing a kinematic fit,
imposing a common vertex on the dimuon and pion tracks,
and constraining the dimuon invariant mass to be the world-average \cPJgy\ mass~\cite{PDG2018}.
The primary vertex (PV) associated with the candidate \Bc is selected among all the reconstructed vertices~\cite{Fruhwirth:2007hz}
as the one with the smallest angle between the reconstructed \Bc momentum and the vector joining the PV with the \Bc decay vertex.
Studies based on simulation show that the probability of selecting a wrong vertex is less than 1\%.
The decay length of the \Bc, denoted by $l$, is computed as the (three-dimensional) distance between the PV and the \Jpsipi vertex
(assumed to be, respectively, the \Bc production and decay vertices).
To avoid biases in the determination of $l$, the PV is refitted without the tracks associated with the muons and the pion.

\begin{figure}[t]
\centering
    \includegraphics[width=\cmsFigWidth]{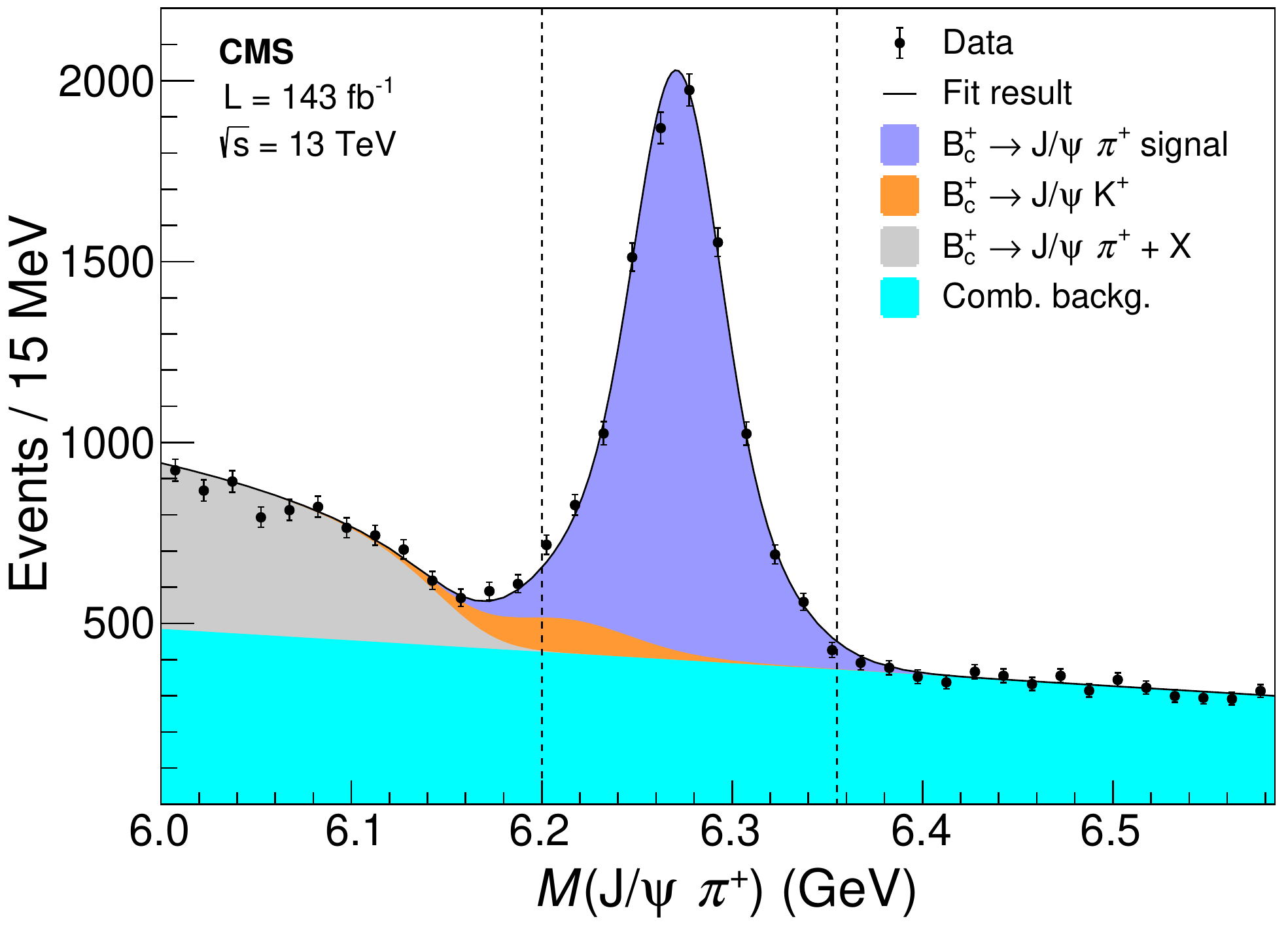}
    \caption{The invariant mass distribution of the \Bc candidates.
    The vertical dashed lines indicate the mass window retained for the reconstruction of the \BcPrime and \BcPrimeStar candidates.
    The vertical bars on the points represent the statistical uncertainty in the data.
    The contributions from various sources are shown by the stacked distributions.
    The solid line represents the result of the fit.}
    \label{fig:BcSignal}
\end{figure}

Similarly to what has been previously done in Refs.~\cite{Khachatryan:2014nfa,Sirunyan:2017nbv},
the \Bc candidates are required to have
$\pt >15$\GeV, rapidity $\abs{y} < 2.4$, $l > 100$\mum, and a kinematic fit $\chi^2$ probability larger than 10\%.
If several \Bc candidates are found in the same event, only the one with the highest \pt is kept.
The invariant mass distribution of the selected \BctoJpsipi candidates, shown in Fig.~\ref{fig:BcSignal},
is fitted to the expected \Bc signal peak, modeled as a sum of two Gaussian functions with a common mean,
superimposed on a background composed of three sources of events:
i)~the combinatorial background resulting from associating the \cPJgy\ with uncorrelated charged particles,
parametrized by a first-order Chebyshev polynomial function;
ii)~partially reconstructed \Bc decays, \BctoJpsiX, only relevant for mass values below 6.2\GeV,
described by a (generalized) ARGUS function~\cite{Albrecht:1990am} convolved with a Gaussian resolution function;
iii)~a small contribution from \BctoJpsiK decays, with a shape determined from simulation studies
and a normalization fixed relative to the \BctoJpsipi yield,
using the ratio of their branching fractions~\cite{lhcb_JpsiK} and the ratio of the reconstruction efficiencies.
The unbinned maximum-likelihood fit gives
a \Bc signal yield of $7629\pm 225$ events,
a \Bc mass of $M(\Bc) = 6271.1 \pm 0.5\MeV$,
and a mass resolution of $33.5 \pm 2.5\MeV$,
where the uncertainties are statistical only.
The measured mass resolution is consistent with the value expected from the simulation studies.
The quality of the fit was evaluated by computing the $\chi^2$ between
the binned distribution and the fit function,
the result being $\chi^2 = 35$ for 30 degrees of freedom.

The \BcPrime and \BcPrimeStar candidates are reconstructed by performing a kinematic fit,
combining a \Bc candidate with two opposite-sign tracks and imposing a common vertex.
Only \Bc candidates with invariant mass in the range 6.2--6.355\GeV are selected.
This mass window, indicated in Fig.~\ref{fig:BcSignal}, reflects the measured \Bc mass and resolution,
with a low-mass edge that, while corresponding to a smaller peak coverage than the high-mass edge,
suppresses the contamination from partially reconstructed decays.
The lifetimes of the \BcPrime and \BcPrimeStar are assumed to be negligible with respect to the measurement resolution,
so that the production and decay vertices essentially overlap.
Therefore, the daughter pions are required to be tracks used in the refitted PV
(a procedure previously followed in Refs.~\cite{cms:Bspi,Sirunyan:2018grk}).
One of the pion candidates must have $\pt > 0.8$\GeV and the other $\pt > 0.6$\GeV.
The \Bcpipi candidates must have $\abs{y} < 2.4$ and a vertex $\chi^2$ probability larger than 10\%.
If several \Bcpipi candidates are found in the same event, only the one with the highest \pt is kept.
Studies with simulated signal samples (providing $S$) and measured sideband events (providing $B$) have shown,
through the $S/\sqrt{S+B}$ figure of merit, that these are optimal event-selection criteria.

\begin{figure}[ht]
\centering
    \includegraphics[width=\cmsFigWidth]{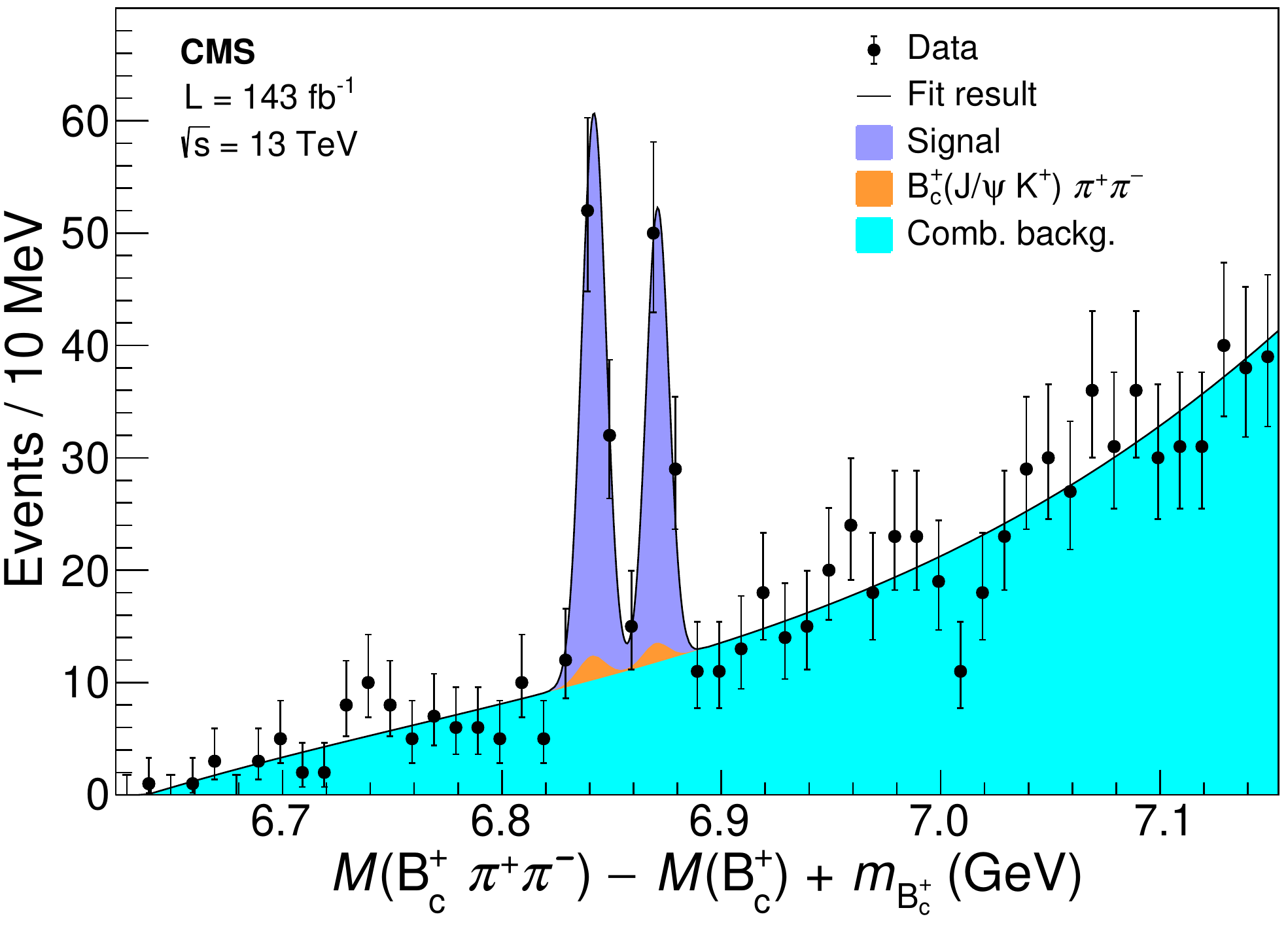}
    \caption{The $M(\protect\Bcpipi) - M(\Bc) + m_{\Bc}$ distribution.
    The \BcPrime is assumed to be the right-most peak.
    The vertical bars on the points represent the statistical uncertainty in the data.
    The contributions from the various sources are shown by the stacked distributions.
    The solid line represents the result of the fit.}
    \label{fig:Bc2sSignal}
\end{figure}

Figure~\ref{fig:Bc2sSignal} shows the $M(\Bcpipi) - M(\Bc) + m_{\Bc}$ distribution, where $M(\Bcpipi)$ and $M(\Bc)$ are, respectively,
the reconstructed invariant masses of the $\Bcpipi$ and \Bc candidates, and $m_{\Bc}$ is the world-average \Bc mass~\cite{PDG2018}.
This variable is measured with a better resolution than $M(\Bcpipi)$ and is, hence, advantageous when searching for peaks in the mass distribution.
The measured distribution is fitted to a superposition of two Gaussian functions,
representing the \BcPrime and \BcPrimeStar signal peaks,
plus a third-order Chebyshev polynomial, modeling the continuum background,
with all parameters left free in the fit.
The two contributions arising from \BctoJpsiK decays are also considered;
they have shapes identical to the signal peaks,
neglecting a shift to lower mass values that should be smaller than 1\MeV,
and normalizations constrained by the ratio of the \BctoJpsiK and \BctoJpsipi signal yields,
as previously mentioned.
The unbinned extended maximum-likelihood fit gives $67 \pm 10$ and $51 \pm 10$
events for the lower-mass and higher-mass peak, respectively.
Since these yields are not corrected for detection efficiencies and acceptances,
they cannot be used to infer ratios of production cross sections.
The two signals are well resolved, their mass difference being
$\Delta M = 29.1 \pm 1.5$\MeV,
where the uncertainty is statistical only.
The widths of the peaks are consistent with the value expected from simulation studies, which is approximately 6\MeV.
The $\chi^2$ between the binned distribution and the fit function is 42 for 39 degrees of freedom.

Studies of simulated samples show that the low-energy photon emitted in the \BcPrimeStar decay
has a very small reconstruction efficiency, of order 1\%.
Consequently, the photon is not detected and the mass of the \BcPrimeStar cannot be measured.
Given the predicted mass splittings mentioned before~\cite{Gregory:2009hq,Dowdall:2012ab,Mathur:2018epb},
the \BcPrimeStar peak is expected to be observed at a
mass lower than the \BcPrime.
The mass of the \BcPrime meson, assumed to be the higher-mass peak in Fig.~\ref{fig:Bc2sSignal}, is measured to be
$6871.0 \pm 1.2$\MeV,
where the uncertainty is statistical only.

The $M(\Bcpipi) - M(\Bc) + m_{\Bc}$ distribution has also been fitted with the two peaks modeled by a Breit--Wigner function,
convolved with a Gaussian resolution function determined from the simulated samples.
The result is that, for both peaks, the natural width parameter of the Breit--Wigner function is consistent with zero,
indicating that both natural widths are small in comparison with the experimental resolution.

The fitting procedure was tested using randomly generated event samples,
of sizes corresponding to the number of measured events,
reflecting the nominal likelihood probability distribution functions and fitted parameters.
No significant fit biases were found in the central values and uncertainties.

Several sources of systematic uncertainties have been considered.
The mass measurements reported here are expected to be essentially insensitive to the event selection criteria.
The analysis was repeated by splitting the data in exclusive sub-samples,
depending on the \Bc rapidity or \pt, or according to the data collection periods.
The \pt thresholds were also varied, between 10 and 18\GeV for the \Bc and
between 3 and 5\GeV for the pion produced in the \Bc decay.
The results remain unchanged, hence no systematic uncertainty is assigned to the selection criteria.
Also, no significant changes are seen in the results when the widths of the Gaussian functions used to describe the two peaks,
or their ratio, are fixed to the values evaluated with the simulated event samples.
The mass measurements might depend on the models used to describe the signal and background contributions.
The impact of the fitting models has been evaluated by varying the considered functional forms.
The combinatorial background, nominally represented by a third-order Chebyshev polynomial,
has been alternatively modeled by the function $(x - x_0)^\lambda \, \exp{(\nu(x - x_0))}$,
where $\lambda$, $\nu$, and $x_0$ are free parameters.
For each of the two signal peaks,
and corresponding \BctoJpsiK terms,
the default Gaussian function was replaced by a Breit--Wigner parametrization.
The differences in the measured observables are taken as the systematic uncertainty associated with the fit modeling.
While the alternative background model leads to a negligible change,
the systematic uncertainties reflecting the modeling of the peaks are 0.8 and 0.7\MeV
in the \BcPrime mass and in $\Delta M$, respectively.

The nominal fit includes a \BctoJpsiK component, with the same shape as the signal peaks and
normalization defined by the expected ratio of the \BctoJpsiK and \BctoJpsipi yields in the \Bc mass window,
corrected by the ratio of the corresponding reconstruction efficiencies.
The normalization has been increased by a factor of two, a variation ten times larger than
the sum of the uncertainties in the ratio of branching fractions~\cite{PDG2018}
and in the ratio of reconstruction efficiencies,
and no significant effect has been seen on the results,
so that no systematic uncertainty is associated with this background contribution.
The \Bc mass distribution includes a contribution from partially reconstructed decays.
Their contamination in the $M(\Bcpipi) - M(\Bc) + m_{\Bc}$ distribution is suppressed by the rejection of \Bc candidates
with invariant mass below 6.2\GeV.
To evaluate possible resolution effects associated with this selection, the requirement was changed to 6.1\GeV,
a variation that also leads to a larger contamination from \BctoJpsiK events.
The difference between the results, taking into account that the two event samples are strongly correlated,
is not statistically significant, so that no systematic uncertainty is assigned.
The potential bias introduced in the mass measurement by possible misalignments of the tracker detectors
has been evaluated through simulation studies and also by comparing distributions measured in the 2016 and 2017 running periods,
a meaningful comparison given that
an important fraction of the CMS tracker detector was replaced between these two years.
The outcome is that the alignment of the detector leads to a negligible systematic uncertainty in the results of the present analysis.
Thus, the total systematic uncertainties are 0.8 and 0.7\MeV,
in the \BcPrime mass measurement and in $\Delta M$, respectively.

The world-average \Bc mass, $m_{\Bc} = 6274.9 \pm 0.8$\MeV~\cite{PDG2018},
enters in the measurement of the \BcPrime mass,
thereby contributing an additional systematic uncertainty of 0.8\MeV.
Strictly speaking, however, it is the mass difference $M(\Bcpipi) - M(\Bc)$ that is measured event by event,
before adding the $m_{\Bc}$ constant, and it is convenient to report the \BcPrime mass as
$M(\BcPrime) - M(\Bc) = 596.1 \pm 1.2 \stat \pm 0.8 \syst$\MeV, a value independent of $m_{\Bc}$.
Another interesting mass difference, also unaffected by the uncertainty in the \Bc world-average mass,
can be derived from the previously reported measurements:
$M(\BcPrimeStar) - M(\BcStar) = [ M(\BcPrime) - M(\Bc) ] - \Delta M = 567.0 \pm 1.0 \, \text{(total)}$\MeV.
Since the systematic effects previously mentioned cancel almost completely in this mass difference,
the total uncertainty is dominated by the statistical term, which was determined by redoing the fit of the
$M(\Bcpipi) - M(\Bc) + m_{\Bc}$ distribution setting this new variable as a floating parameter,
to properly account for the correlations between the parameters.
The observation of two peaks, rather than one, is established with a significance of 6.5 standard deviations,
evaluated with the likelihood-ratio technique confronting the two-peaks (ten free parameters)
and one-peak (seven free parameters) hypotheses,
using asymptotic formulae~\cite{Wilks:1938dza,Cowan:2010js}
and accounting for the (dominant) systematic uncertainty in the signal model.

In summary, signals consistent with the \BcPrime and \BcPrimeStar states have been separately observed for the first time
by investigating the \Bcpipi invariant mass spectrum measured by CMS.
The analysis is based on the entire LHC sample of proton-proton collisions at a center-of-mass energy of 13\TeV,
corresponding to a total integrated luminosity of 143\fbinv.
The two peaks are well resolved, with a measured mass difference of
$\Delta M = 29.1 \pm 1.5\stat \pm 0.7\syst$\MeV.
The \BcPrime mass is measured to be
$6871.0 \pm 1.2 \stat \pm 0.8\syst \pm 0.8 \, (\Bc)$\MeV,
where the last term is the uncertainty in the world-average \Bc mass.
Because the low-energy photon emitted in the $\BcStar \! \to \! \Bc \, \gamma$ radiative decay is not reconstructed,
the observed \BcPrimeStar peak has a mass lower than the true value, which remains unknown.
These measurements contribute significantly to the detailed characterization of heavy meson spectroscopy
and provide a rich source of information on the nonperturbative QCD processes that bind heavy quarks into hadrons.

\begin{acknowledgments}

 We congratulate our colleagues in the CERN accelerator departments for the excellent performance of the LHC and thank the technical and administrative staffs at CERN and at other CMS institutes for their contributions to the success of the CMS effort. In addition, we gratefully acknowledge the computing centers and personnel of the Worldwide LHC Computing Grid for delivering so effectively the computing infrastructure essential to our analyses. Finally, we acknowledge the enduring support for the construction and operation of the LHC and the CMS detector provided by the following funding agencies: BMBWF and FWF (Austria); FNRS and FWO (Belgium); CNPq, CAPES, FAPERJ, FAPERGS, and FAPESP (Brazil); MES (Bulgaria); CERN; CAS, MoST, and NSFC (China); COLCIENCIAS (Colombia); MSES and CSF (Croatia); RPF (Cyprus); SENESCYT (Ecuador); MoER, ERC IUT, and ERDF (Estonia); Academy of Finland, MEC, and HIP (Finland); CEA and CNRS/IN2P3 (France); BMBF, DFG, and HGF (Germany); GSRT (Greece); NKFIA (Hungary); DAE and DST (India); IPM (Iran); SFI (Ireland); INFN (Italy); MSIP and NRF (Republic of Korea); MES (Latvia); LAS (Lithuania); MOE and UM (Malaysia); BUAP, CINVESTAV, CONACYT, LNS, SEP, and UASLP-FAI (Mexico); MOS (Montenegro); MBIE (New Zealand); PAEC (Pakistan); MSHE and NSC (Poland); FCT (Portugal); JINR (Dubna); MON, RosAtom, RAS, RFBR, and NRC KI (Russia); MESTD (Serbia); SEIDI, CPAN, PCTI, and FEDER (Spain); MOSTR (Sri Lanka); Swiss Funding Agencies (Switzerland); MST (Taipei); ThEPCenter, IPST, STAR, and NSTDA (Thailand); TUBITAK and TAEK (Turkey); NASU and SFFR (Ukraine); STFC (United Kingdom); DOE and NSF (USA).

\end{acknowledgments}

\bibliography{auto_generated}
\cleardoublepage \appendix\section{The CMS Collaboration \label{app:collab}}\begin{sloppypar}\hyphenpenalty=5000\widowpenalty=500\clubpenalty=5000\vskip\cmsinstskip
\textbf{Yerevan Physics Institute, Yerevan, Armenia}\\*[0pt]
A.M.~Sirunyan$^{\textrm{\dag}}$, A.~Tumasyan
\vskip\cmsinstskip
\textbf{Institut f\"{u}r Hochenergiephysik, Wien, Austria}\\*[0pt]
W.~Adam, F.~Ambrogi, T.~Bergauer, J.~Brandstetter, M.~Dragicevic, J.~Er\"{o}, A.~Escalante~Del~Valle, M.~Flechl, R.~Fr\"{u}hwirth\cmsAuthorMark{1}, M.~Jeitler\cmsAuthorMark{1}, N.~Krammer, I.~Kr\"{a}tschmer, D.~Liko, T.~Madlener, I.~Mikulec, N.~Rad, J.~Schieck\cmsAuthorMark{1}, R.~Sch\"{o}fbeck, M.~Spanring, D.~Spitzbart, W.~Waltenberger, J.~Wittmann, C.-E.~Wulz\cmsAuthorMark{1}, M.~Zarucki
% ...
\vskip\cmsinstskip
\dag: Deceased\\
1:  Also at Vienna University of Technology, Vienna, Austria\\
% ...
\end{sloppypar}
\end{document}